\documentclass[preprint]{elsarticle}

\usepackage{verbatim} 
\usepackage{threeparttable}
\usepackage{graphicx}        
\begin{document}

\title{Synthesis of neutron-rich transuranic nuclei in fissile spallation targets}

\author[fias,kurch]{Igor Mishustin}
\ead{mishustin@fias.uni-frankfurt.de}
\author[fias,inr]{Yury Malyshkin}
\ead{malyshkin@fias.uni-frankfurt.de}
\author[fias,inr]{Igor Pshenichnov}
\ead{pshenich@fias.uni-frankfurt.de}
\author[fias]{Walter Greiner}

\address[fias]{Frankfurt Institute for Advanced Studies, J.-W. Goethe University, 60438 Frankfurt am Main, Germany}
\address[kurch]{``Kurchatov Institute'', National Research Center, 123182 Moscow, Russia}
\address[inr]{Institute for Nuclear Research, Russian Academy of Science, 117312 Moscow, Russia}

\begin{abstract}

A possibility of synthesizing neutron-reach super-heavy elements in spallation 
targets of Accelerator Driven Systems (ADS) is considered.
A dedicated software called Nuclide Composition Dynamics (NuCoD) was developed to model
the evolution of isotope composition in the targets during a long-time irradiation 
by intense proton and deuteron beams. Simulation results show that transuranic 
elements up to $^{249}$Bk can be produced in multiple neutron capture reactions
in macroscopic quantities. However, the neutron flux achievable in a spallation
target is still insufficient to overcome the so-called fermium gap. Further 
optimization of the target design, in particular, by including moderating 
material and covering it by a reflector will turn ADS into an alternative 
source of transuranic elements in addition to nuclear fission reactors. 
\end{abstract}

\maketitle

\section{Introduction}
\label{sec:intro}

Neutrons propagating in a medium induce different types of nuclear reactions 
depending on their energy. Apart of the elastic scattering, the main reaction 
types for low-energy neutrons are fission and neutron capture, which dominate, 
respectively, at higher and lower energies with the boarder line around 1~MeV. 
Generally, the average number of neutrons captured by a nucleus $A$ during a  
time interval of $\Delta t$ is given by the formula:
\begin{equation}
  \Delta N = f ~\sigma_{nA}~ \Delta t~,
\end{equation}
where $\sigma_{nA}=(1\div3)\: \mathrm{b}$ is the 
cross section of (n,$\gamma$) reaction on heavy nuclei and
 $f=\langle v\frac{\mathrm d n}{\mathrm d v} \rangle $ is the neutron 
flux averaged over the velocity distribution. 

The possibility to capture many neutrons in nuclear explosions was 
considered already in 60s and 70s, see e.g. ref.~\cite{Bell1967}, and recently 
in refs.~\cite{Botvina2010,Zagrebaev2011a}. In this case the neutron flux is 
about $3\cdot10^{30} {\rm \frac{n}{s \cdot cm^2}}$, and  the explosion time 
$\Delta t \simeq 1$~$\mu$s, therefore:
  $$\Delta N_{expl} = f ~ \Delta t ~ \sigma = 
    3 \cdot 10^{30} \mathrm{\frac{n}{s \cdot cm^2}} \cdot 10^{-6} \mathrm s
    \cdot 10^{-24} \mathrm{cm^2} \simeq 3, $$
assuming that $\sigma_{nA} \simeq 1$~b. As shown in~\cite{Zagrebaev2011a}, 
macroscopic quantities of superheavy (SH) elements located on the Island of 
Stability can be produced in multiple nuclear explosions.

In ordinary nuclear fission reactors the average neutron flux is typically below 
10$^{15}$ $\mathrm{n/(s \cdot cm^2)}$, and the nuclei up to fermium ($Z$=100) can be 
produced there~\cite{Seaborg1968}. In this paper we consider the 
possibility of producing neutron-rich transuranic (TRU) nuclei in Accelerator Driven 
Systems (ADS) with fissile spallation targets made of americium ($Z$=95). In this case the 
average neutron flux could be as large as 
$3 \cdot 10^{16} \mathrm{n/(s \cdot cm^2)}$~\cite{Malyshkin2014}. 
Since the (n,$\gamma$) reaction is only efficient at relatively low energies, 
$E_n<1$~MeV, the flux of such neutrons is about twice as low.
However, the target can be irradiated by a proton or deuteron beam for a long time, 
e.g. for one year. Then taking $\sigma_{nA}=2$~b, one obtains:
  $$ \Delta N_{ADS} ~\simeq 3 \cdot 10^{16} \mathrm{\frac{n}{s \cdot cm^2}}
    \cdot 3 \cdot 10^7 \mathrm s \cdot 2 \cdot 10^{-24} \mathrm{cm^2} 
    \simeq 2 $$~.

The distribution of nuclides in the number of captured neutrons follows a Poisson
distribution, providing that all other reactions are neglected, see, e.g.~\cite{Botvina2010}.
With a simple-minded correction for losses due to fission it can be written as:
\begin{equation}
\label{eq:poisson}
 P_n(\bar n) = \frac{\bar n^n e^{- \bar n}}{n!} (1-q_{fission})^n
\end{equation}
where $q_{fission} \simeq \frac{1}{2}$ is the probability of fission in each 
step. With $\bar n = 2$ and $q_{fission}=\frac{1}{2}$ the probability of 
capture of 20 neutrons, to produce isotopes with atomic mass number $A+20$, can be 
estimated as $6 \cdot 10^{-20}$. For a small Am target with its mass of about 40~kg 
the number of nuclei in the target is $\sim 10^{26}$, therefore, $\sim 10^6$ 
nuclei with mass A+20 can be synthesized after one year of operation.

Neutrons suitable for the production of super-heavy elements should have softer spectrum to increase 
the efficiency of neutron capture reactions. Therefore, such targets should include 
some moderating material to slow down neutrons. However, too intensive moderation of 
neutrons would suppress the neutron breeding in fission reactions. A special investigation is 
required to find a compromise solution. In this paper we model an Am target equipped with
a Be reflector by performing detailed Monte Carlo simulations of nuclear reactions 
and using nuclear data files with realistic neutron interaction cross sections.

\section{Modelling of fissile spallation targets}
\label{sec:simulation}

\subsection{MCADS --- Monte Carlo model for Accelerator Driven Systems}
\label{ssec:MCADS}

Simulations in this work are performed with the Monte Carlo model for 
Accelerator Driven Systems (MCADS), which is based on the Geant4 toolkit 
(version 9.4 with patch 01)~\cite{Agostinelli2003,Allison2006,Apostolakis2009}. 
This model was used for modeling neutron production and transport in 
spallation targets made of tungsten, uranium and americium~\cite{Malyshkin2012a,Malyshkin2014}.
MCADS is capable to visualize target volumes as well as histories of primary protons and 
all secondary particles. It has flexible scoring techniques to calculate neutron flux, 
heat deposition inside the target and leakage of particles from the target.
MCADS makes possible to employ several cascade models for the fast initial stage 
of pA and dA interactions, which are combined with evaporation, multi-fragmentation and Fermi break-up 
models for the slow de-excitation stage of thermalized residual nuclei. 

MCADS was validated with three different cascade models: Bertini Cascade and Binary 
Cascade coupled with their standard fission-evaporation codes and Intra-Nuclear 
Cascade Li\`ege (INCL) coupled with the ABLA model. 
Calculations were performed for thin and thick targets made of tungsten and 
uranium~\cite{Malyshkin2012a}. For thin $^{238}$U targets irradiated by 27, 63 and 
1000~MeV protons we have analyzed data on the fission cross sections, neutron 
multiplicity and mass distribution of fission products, and the best agreement 
was obtained for calculations involving INCL/ABLA. In particular, as demonstrated
in ref.~\cite{Malyshkin2012a} MCADS results obtained 
with INCL/ABLA model are in good agreement with experimental data available 
for extended tungsten and uranium targets irradiated by protons in the energy range 
of $400-1470$~MeV.

Several extensions of the Geant4 source code and additional nuclear 
data~\cite{Mendoza2012} are needed for the description of 
proton- and neutron-induced reactions and elastic scattering on 
TRU nuclei~\cite{Malyshkin2013a}. MCADS with the 
modified Geant4 was validated with $^{241}$Am and $^{243}$Am thin targets.  
A good agreement with proton- and neutron-induced fission 
cross sections, fission neutron spectra, neutron multiplicities, fragment mass 
distributions and neutron capture cross sections measured in experiments 
was demonstrated~\cite{Malyshkin2013}.

\subsection{Criticality studies}

\begin{figure}[htb]
\center \includegraphics[width=0.8\columnwidth]{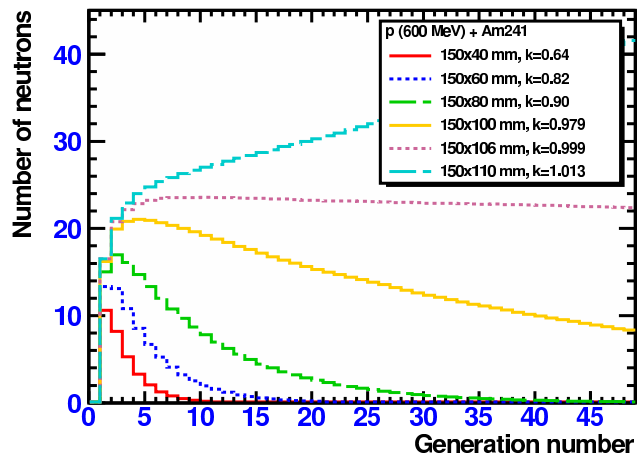}
\caption{Number of neutrons $N_i$ in $^{241}$Am targets of 15~cm length and 
radii of 40, 60, 80, 100, 106 and 110~mm as a function of neutron generation number $i$.}
\label{fig:Criticality}
\end{figure}

When dealing with fissile materials one must control the criticality of the 
target. This can be done by calculating the neutron multiplication 
factor $k = \langle {N_i}/{N_{i-1}} \rangle$, which is defined as 
the average ratio between the numbers of neutrons, $N_{i-1}$ and ${N_i}$, 
in two subsequent generations $i-1$ and $i$. 
Obviously, the target mass should not exceed the so-called critical mass which corresponds to the 
condition $k=1$. The calculated number of neutrons in the first 50 neutron
generations developing in a cylindrical $^{241}$Am target following an impact
of a beam proton is shown in Figure~\ref{fig:Criticality}.
The results are given for a set of targets of the same length (150~mm), but 
different radii in the range from 40 to 110~mm. As one can see, $k \simeq1$ in the target with the 
radius of 106~mm, and the number of neutrons in subsequent generations remains constant. 
A supercritical regime ($k>1$) is reached for larger targets where the number of 
neutrons grows from one generation to another.

The main safety requirement to a spallation target made of a fissile 
material is that it is deeply subcritical and operates in a safe mode.
Cylindrical Am targets, which are characterized by the neutron 
multiplication factor $k \le 0.6$, are considered below.

\section{Changes in isotope composition of spallation targets during irradiation}

\subsection{NuCoD -- Nuclide Composition Dynamics}
\label{ssec:NuCoD}

A significant amount of nuclides can be produced and accumulated in a 
spallation target during its long-term irradiation. Moreover, nuclides
of interest can be produced specifically in nuclear reactions on those nuclides,
which were absent in the original target material before the beginning of irradiation.  
In this respect the MCADS model has a disadvantage of neglecting reactions induced
on new nuclides produced in the target during irradiation. A burn-up of the 
target material is also neglected. This stems from a restriction
of the Geant4 toolkit which is incapable of changing the definitions of materials used 
to build geometry objects until a required number of events is obtained, i.e. during a
simulation run. This restriction does not impair the accuracy of MCADS results for 
a short-term irradiations, when such changes can be neglected.    

The Nuclide Composition Dynamics (NuCoD) code was developed on the basis of MCADS 
for modeling long-term irradiation of spallation targets taking into account the 
above-described changes of their nuclide composition during the irradiation period.
Nuclides produced in spallation targets can be transmuted into other nuclei 
in reactions induced by beam projectiles and various secondary particles. Furthermore, 
unstable isotopes undergo radioactive decays. The algorithm of modeling consists of 
several steps:
\begin{enumerate}
 \item Monte Carlo simulation with a target of a given composition to calculate 
the yields of products of spallation, fission and neutron capture reactions;  
 \item calculation of changes of such yields in time due to respective radioactive decays;
 \item update of the target composition and preparing input for the next 
Monte Carlo simulation. 
\end{enumerate}

Following this simulation scheme the Geant4-based code MCADS described 
above in Section~\ref{ssec:MCADS} is used for Monte Carlo simulations 
with targets of constant nuclide composition. MCADS is also used to verify that 
each target composition is subcritical. In turn, NuCoD  
uses the yields of newly produced and burned-up nuclides calculated by 
MCADS at a previous step to estimate the isotope abundances in the target for a given 
beam intensity and exposure time. The key feature of MCADS consists in the ability to 
simulate targets containing minor actinides (MA). This makes possible to  
calculate the yields of TRU elements produced in a long chain 
of nuclear reactions and radioactive decays in targets containing MA by using NuCoD.

It is not reasonable to run MCADS for the targets containing all the isotopes 
produced at previous steps. Firstly, it is too expensive in terms of 
CPU time, as simulations with multicomponent materials require more 
computational steps. Secondly, many isotopes in the target do not 
influence significantly the yields of the isotopes of our interest and thus may be 
omitted. Thirdly, the abundances of some isotopes may be too low. As a result, 
the number of reactions generated in an MCADS run on such rare nuclei may be insufficient 
for reliable estimations of the yields of respective reaction products. 

In order to solve all these problems, a special technique was implemented.
Instead of modeling a target containing all the isotopes with their actual 
abundances, a set of special targets is considered at each simulation step $n$. 
The first one, called the base target, contains only the most abundant isotopes. 
Then, separate modified targets, referred to as isotope-enriched targets, 
are constructed for each isotope $(A_i, Z_i)$ with its abundance $a(A_i,Z_i,n)$ 
less than $a_{th1}=0.005$, but which affects significantly the final yields of 
isotopes we are interested in. These modified targets are constructed as follows:
\begin{itemize}
\item The base target contains only isotopes with abundances exceeding a 
threshold value $a_{th0}$ (by default $a_{th0}=0.001$). Additionally, the 
isotopes, for which the isotope-enriched targets are built, are excluded from 
the base target.
\item In the $i$-th isotope-enriched target the concentration of the isotope 
$i$ is enhanced artificially. Its abundance is set to $a_{th1}=0.005$ and the 
corresponding scale factor $k_i(n)$ is stored. 
\end{itemize}
Running MCADS with these targets makes possible to calculate the amounts of 
newly produced and consumed isotopes at each time step $\Delta t$. 
$\Delta \tilde{a}(A,Z,n)$ is defined as the abundance variation of an isotope 
with mass $A$ and charge $Z$ calculated with MCADS for the base target during 
the step $n$. Analogously, $\Delta a_i(A,Z,n)$ is calculated for each 
isotope-enriched target. Finally, the abundance of the isotope with mass $A$ and 
charge $Z$ at the beginning of the step $(n+1)$ is calculated as:
\begin{equation}
\label{eq:a_n+1}
a(A,Z,n+1) = a(A,Z,n) + \Delta \tilde{a}(A,Z,n) + \\
  \sum_{i} \frac{1}{k_i(n)} (\Delta a_i(A,Z,n) - \Delta \tilde{a}(A,Z,n)).
\end{equation}

Finally, the evolution of the target composition due to decays is calculated 
using a newly developed recursive algorithm, which in principle can be 
considered as an extension of Bateman equations~\cite{Cetnar2006} for 
the case of a continuously irradiated target. This algorithm is described in Appendix A.

\subsection{NuCoD validation}
\label{ssec:NuCoD_validation}

\begin{figure}[!htb]
\begin{centering}
\includegraphics[width=0.85\columnwidth]{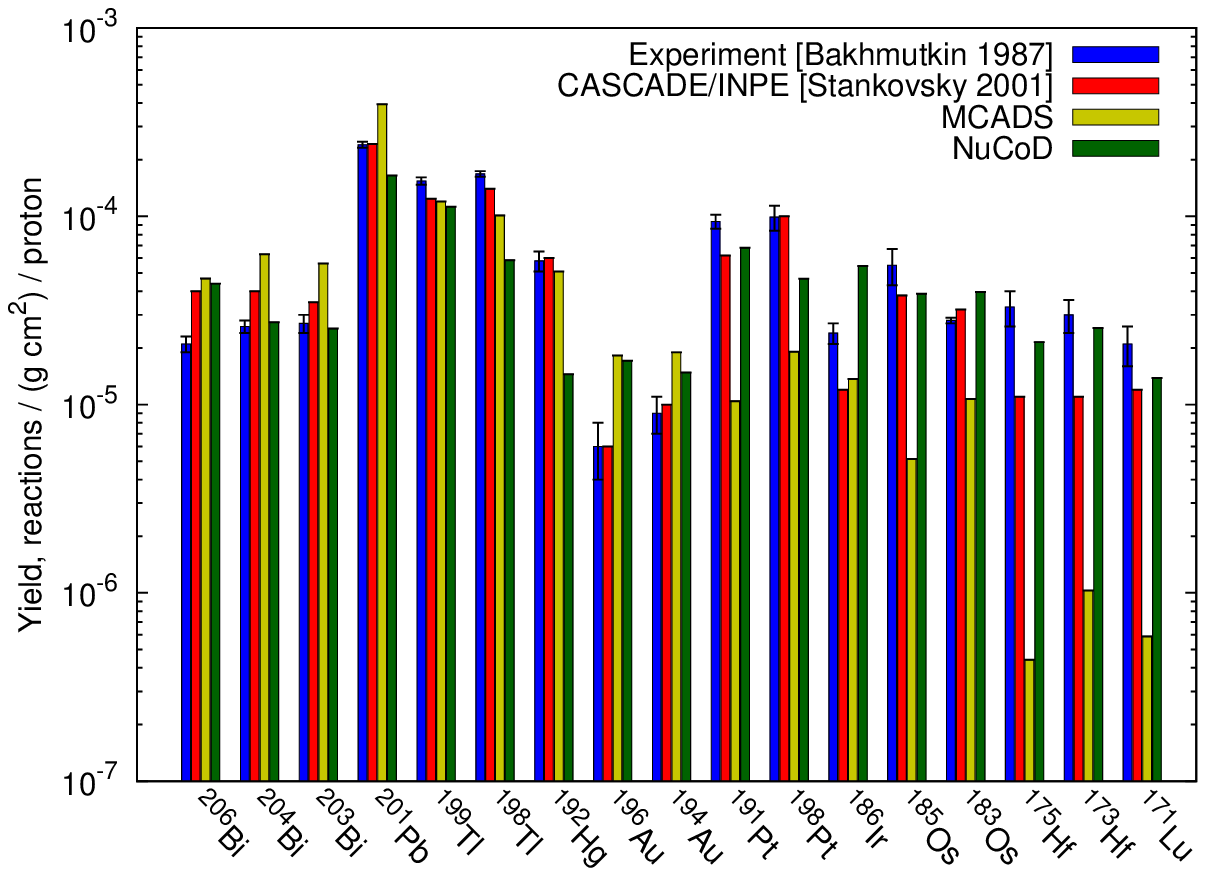}
\includegraphics[width=0.85\columnwidth]{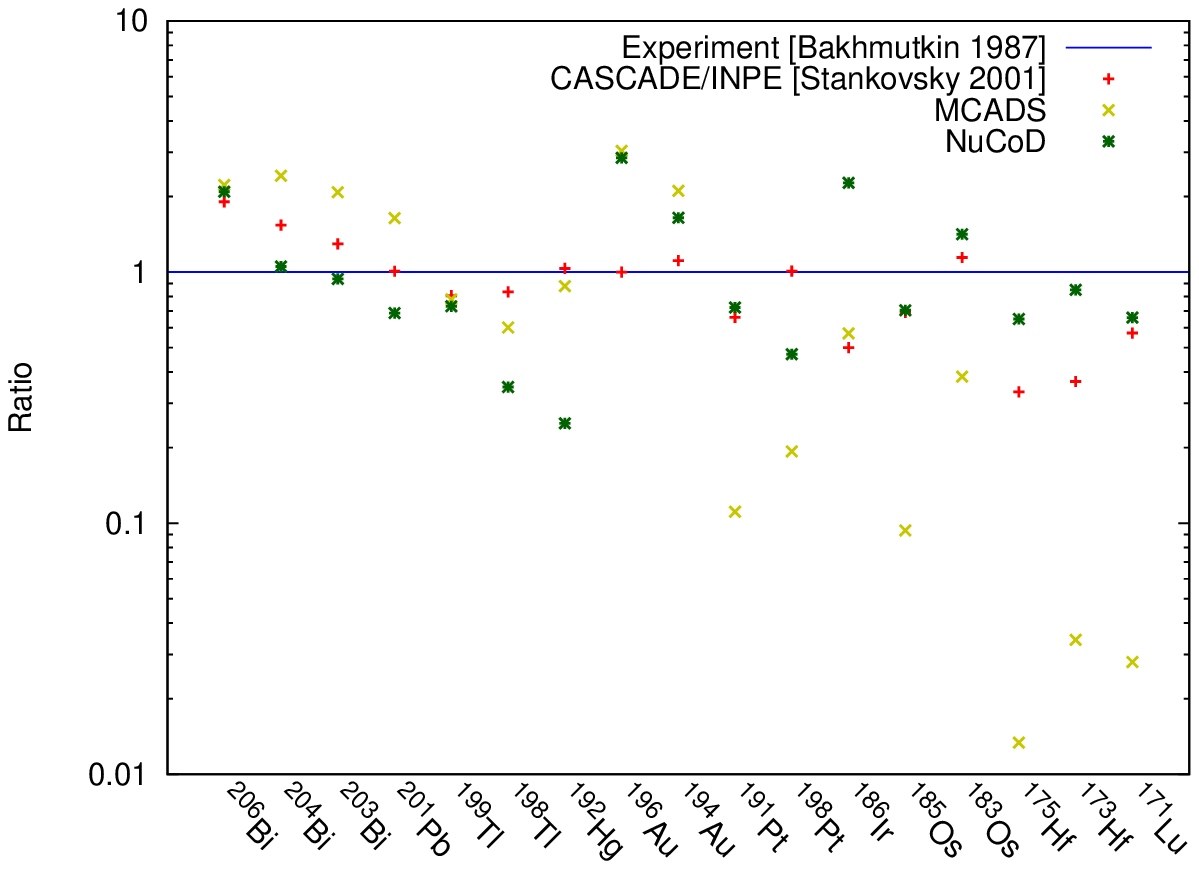}
\caption[Yields of spallation products in experiment~\cite{Bakhmutkin1986}
compared with results of computations]
{Yields of spallation products in the lead target irradiated by 1~GeV protons
(top panel). Experimental data are from ref.~\cite{Bakhmutkin1986}. 
Results of calculations with CASCADE/INPE~\cite{Stankovsky2001}, MCADS and NuCoD 
are presented by different colors explained in the legend. The calculated yields divided
by the measured yields (bottom panel), see the legend for notations.}
\label{fig:spal_prod}
\end{centering}
\end{figure}

In order to validate NuCoD an irradiation of a lead target by protons 
as described in~\cite{Bakhmutkin1986} was simulated. In the 
experiment~\cite{Bakhmutkin1986} a target with the diameter of 20~cm and 
length of 60~cm was irradiated by a 1~GeV proton beam with a current of 
0.3~$\rm \mu$A for 3~hours. The measurements of $\gamma$-spectra started
7~hours after the termination of the irradiation and lasted next 8~days. 
In Figure~\ref{fig:spal_prod} the yields of spallation products measured at 
the depth of 6~cm are compared with results of NuCoD simulation. Results of calculations 
obtained only with MCADS, i.e. neglecting decays of spallation products, are also
shown in Figure~\ref{fig:spal_prod} together with the results of 
\mbox{CASCADE/INPE} code reported in ref.~\cite{Stankovsky2001}. 

As seen from Figure~\ref{fig:spal_prod}, the results of both codes, 
\mbox{CASCADE/INPE} and MCADS, deviate from the measured yields. However, 
\mbox{CASCADE/INPE} results are much closer to the data compared to MCADS, which 
calculates prompt yields and neglects radioactive decays of 
spallation products. In contrast, NuCoD simulates isotope production in the 
target and also respective decay chains. This notably improves the agreement with
the measured yields. This is especially important for the domain of lighter 
elements (from Pt to Lu), where their abundances are significantly 
increased due to decay products of parent nuclides.

\section{Isotope production in a Am(OH)$_3$ target}

\subsection{Geometry and composition of the target}

The yields of heavy transuranic elements, up to fermium, were calculated with 
NuCoD in a cylindrical target irradiated by a 1~GeV deuteron beam with a current 
of 20~mA. The advantage of deuterons over protons with respect to neutron production in
spallation targets was demonstrated in ref.~\cite{Hashemi-Nezhad2011}. This 
motivated our choice of projectiles in the present work. We consider a target 
of the 6~cm radius and 36~cm length made of Am(OH)$_3$. 
Due to a lack of information on the americium hydroxide properties its
density was set to 10~g/cm$^3$. With this density of the target material 
the mass of the target was estimated as 40.7~kg. 

\begin{figure}[!htb]
\begin{centering}
\includegraphics[height=0.28\textheight]{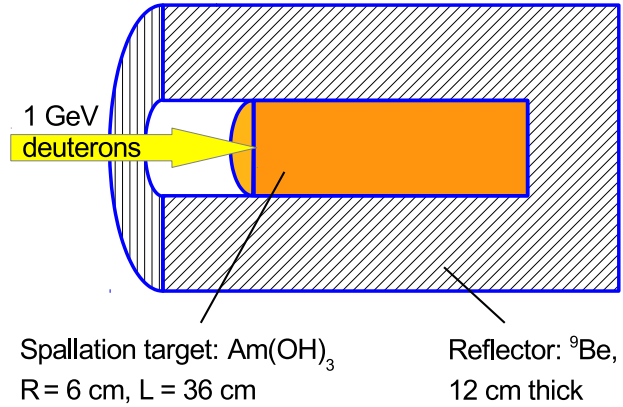}
\caption[Spallation target for production of SH nuclei]
{Layout of a spallation target for production of SH nuclei.}
\label{fig:SHE_target}
\end{centering}
\end{figure}

Since the masses of neutrons and protons are almost equal, fast neutrons 
transfer to protons a half of their energy on average in their elastic scattering
on hydrogen nuclei. Thus hydrogen containing in Am(OH)$_3$ moderates neutrons 
very efficiently. The probability of capture increases with the decrease of 
neutron energy and the production rate of neutron-rich elements also increases.
In order to prevent, at least partially, leakage of neutrons, it was assumed that 
the target was covered by a 12~cm thick beryllium layer as shown in 
Figure~\ref{fig:SHE_target}. Since beryllium has a low 
neutron capture cross section, this material is widely used for reflecting
neutrons in nuclear reactors.

The isotope ratio of americium was taken approximately as in spent nuclear fuel, 
\mbox{$^{241}$Am $:$ $^{243}$Am} $= 3:2$.  Since the probability to 
capture a certain number of neutrons drops rapidly with the number of 
required capture events, the presence of $^{243}$Am in the target enhances the 
production of neutron-rich SH nuclei due to two additional neutrons
containing in $^{243}$Am with respect to $^{241}$Am. 
According to our Mote Carlo simulation results, the average number of neutrons 
captured by $^{243}$Am or heavier nuclides during 6 months of irradiation is about 
$\bar{n}(0.5 \rm y)=0.6$ and $\bar{n}(1 \rm y)=1.2$ during one year of irradiation. 
This is a more accurate result compared with the estimation given above 
in Section~\ref{sec:intro}.

\subsection{Dynamical nuclear charts}

\begin{figure}[!htb]
\begin{centering}
{\includegraphics[width=0.49\textwidth]{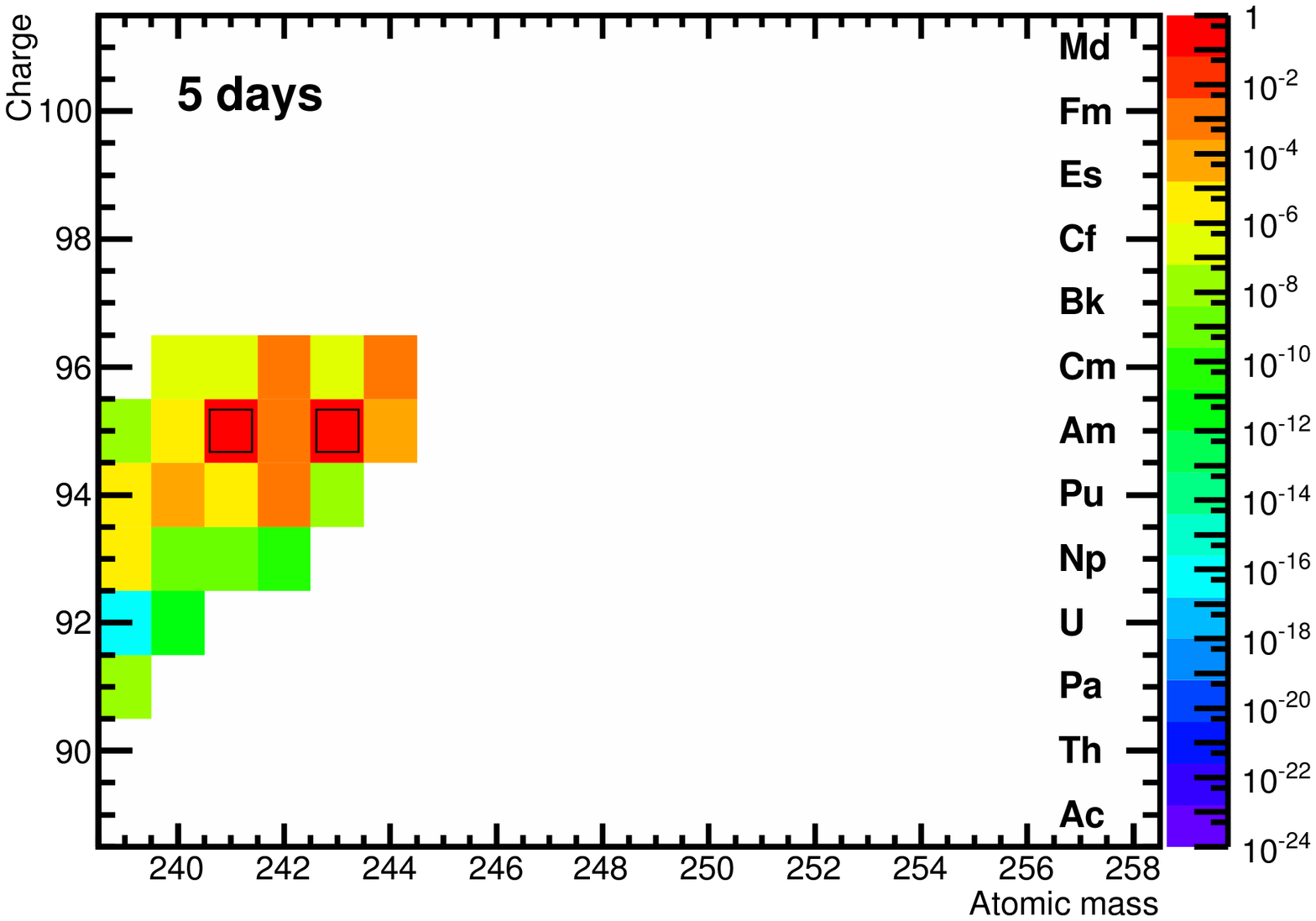}}
{\includegraphics[width=0.49\textwidth]{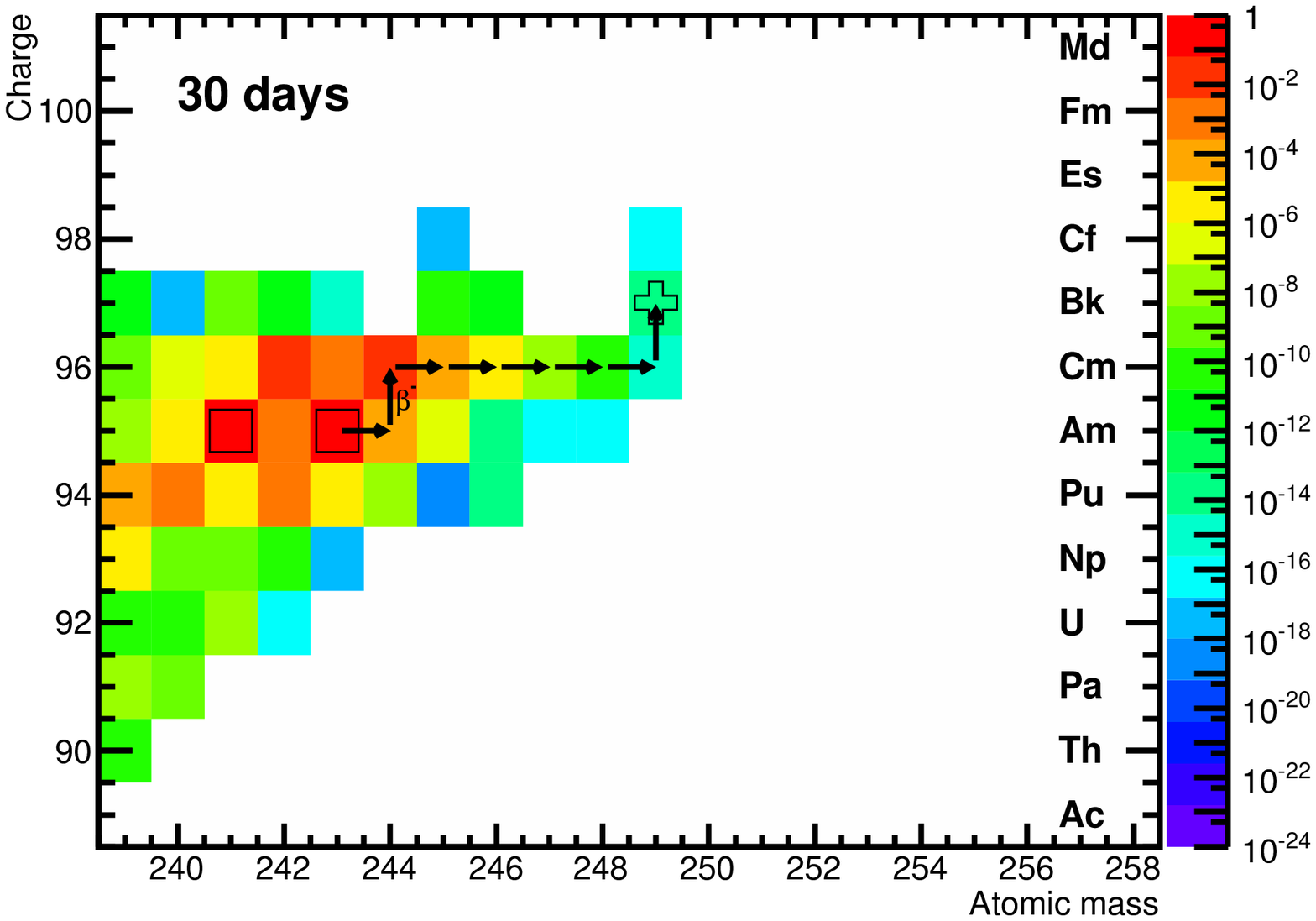}}
{\includegraphics[width=0.49\textwidth]{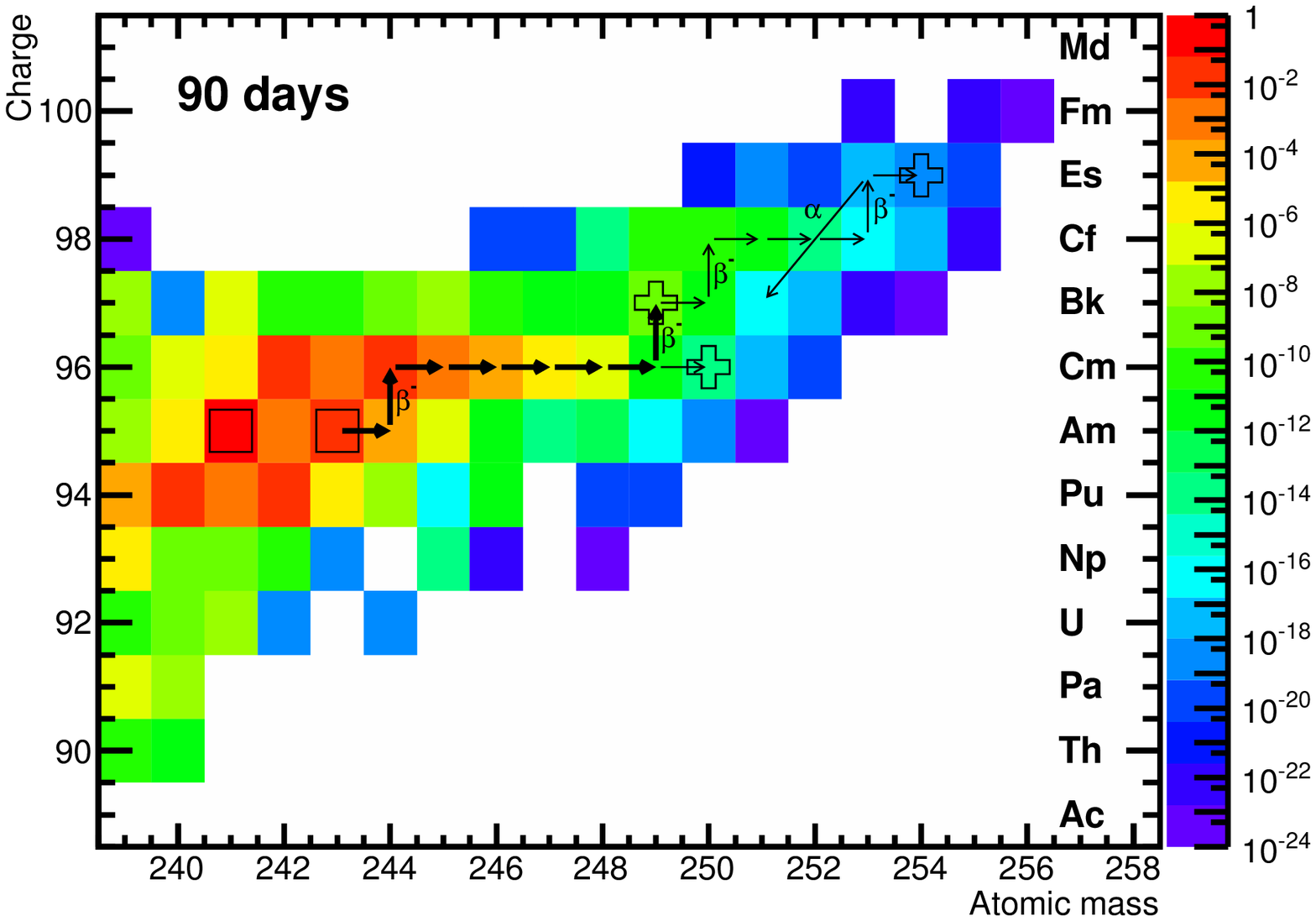}}
{\includegraphics[width=0.49\textwidth]{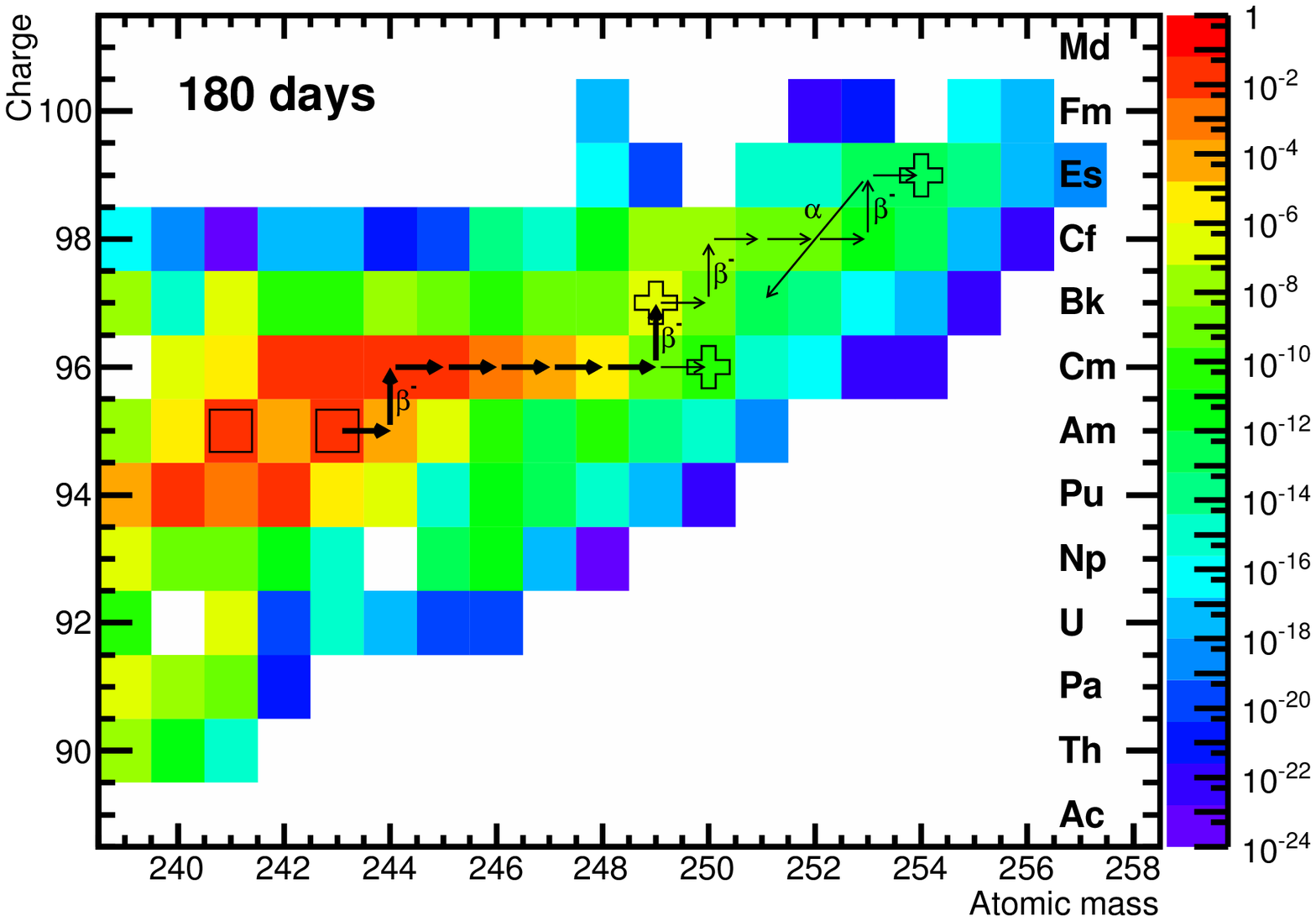}} 
\caption[Maps of isotopes in the Am(OH)$_3$ target with reflector]
{Calculated maps of isotope abundance after 5, 30, 90 and 180~days of irradiation. 
The initial isotopes, $^{241}$Am and $^{243}$Am, are marked with the 
squares, the crosses indicate $^{249}$Bk, $^{250}$Cm and $^{254}$Es.}
\label{fig:iso_maps}
\end{centering}
\end{figure}

Below we present our results obtained with the time step of 5~days between the target 
composition updates. The maps of isotopes contained in the target after 5, 30, 90 
and 180~days of irradiation are presented in Figure~\ref{fig:iso_maps}. 
The first map represents the isotopes produced before the first target 
composition update, i.e.\ a direct 
output of MCADS simulation corrected for the radioactive decays. Therefore, 
only products of (d,Am), (p,Am) and (n,Am) reactions and their decay products are present 
on this map. 

The main path to produce TRU elements (starting from $^{243}$Am) is 
shown by thick arrows extending up to $^{249}$Bk. $^{244}$Am is produced in 
neutron capture reaction, and it undergoes $\beta^-$-decay with the half-life 
time of $t_{1/2}=10$~hours. A resulting long-lived isotope $^{244}$Cm 
($t_{1/2}=18$~years) can capture more neutrons. The next four isotopes of Cm 
have half-life times of several thousands years and longer and can be 
accumulated in the target after corresponding number of $(n,\gamma)$ reactions. 
The isotope $^{249}$Cm has the half-life time of 64~minutes and decays into a relatively 
long-lived $^{249}$Bk ($t_{1/2}=330$~days) which we are interested in. 
Heavier isotopes are produced in similar processes depicted by thin 
arrows in Figure~\ref{fig:iso_maps} which extend up to 
fermium. The isotopes $^{254-256}$Fm have half-life times of 2.5--20~hours and 
undergo $\alpha$-decay or spontaneous fission and drop out from the chain of 
subsequent neutron capture reactions. That is why this specific domain in the map
of isotopes is called the fermium gap. Due to its presence a further move to 
heavier isotopes is practically impossible. Indeed, as seen from 
Figure~\ref{fig:iso_maps}, the abundance of Fm isotopes in the considered 
spallation targets is extremely low: 
$\sim 10^{-17}$.

The production of $^{250}$Cm and $^{254}$Es is suppressed, respectively, by a prompt  
$\beta^-$-decay of $^{249}$Cm and the $\alpha$-decay of $^{253}$Es. However, the 
situation may change in the case of a pulse-mode accelerator with a short pulse duration 
of $\sim 1$~$\rm \mu$s and very intense bunches of protons.

\subsection{Abundances of produced nuclides}

\begin{figure}[!htb]
\begin{centering}
\includegraphics[width=0.85\columnwidth]{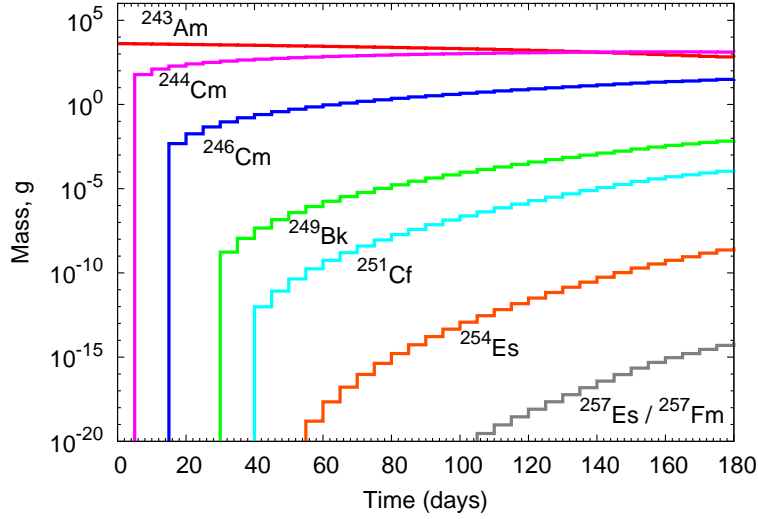}
\caption[Time evolution of some isotopes abundances in the Am(OH)$_3$ 
target with reflector]
{Time evolution of the amount of $^{243}$Am, $^{244}$Cm, $^{246}$Cm, 
$^{249}$Bk, $^{251}$Cf, $^{254}$Es and $^{257}$Es/$^{257}$Fm calculated with NuCoD 
for the Am(OH)$_3$ target irradiated by 1~GeV deuteron beam.}
\label{fig:te}
\end{centering}
\end{figure}

The time evolution of the abundances of $^{243}$Am, $^{244,246}$Cm, $^{249}$Bk, 
$^{251}$Cf, $^{254}$Es and $^{257}$Es/$^{257}$Fm 
is shown in Figure~\ref{fig:te}. As one can see, the abundance 
of $^{244}$Cm raises rapidly at the beginning and after $\sim 135$~days exceeds 
the $^{243}$Am abundance. In order to produce $^{246}$Cm three neutrons must be 
captured by a single nucleus. This means that $^{246}$Cm can be obtained only after 
three simulation steps of NuCoD, i.e. after 15~days. This deficiency of 
the simulation method is noticeable only at the first steps and much less affects the 
results in further steps. As seen from Figure~\ref{fig:te}, the amount of $^{249}$Bk 
reaches the level of 10~mg after 150~days.

The abundance of selected long-lived isotopes after 2, 4 and 6~months of irradiation 
are presented in Table~\ref{table:SHE_masses}. One can see that the amount of $^{249}$Bk 
exceeds the level of several milligrams, which is sufficient to fabricate a fusion 
target~\cite{Zagrebaev2012} to be used in experiments on production of SHE 
elements in nucleus-nucleus collisions. Due to the 
$\beta^-$-decay of $^{249}$Bk ($t_{1/2}=320$~days) $^{249}$Cf can also reach 
this level after some cooling time after the end of irradiation by deuterons. 
\begin{table}[!htb]
\caption{Abundance of $^{249}$Bk, $^{249,251}$Cf, $^{250}$Cm, $^{254}$Es and 
$^{257}$Es/$^{257}$Fm after 2, 4 and 6~months of irradiation by a 1~GeV deuteron beam with 
a current of 20~mA.}
\centering
\begin{threeparttable}
\begin{tabular}{cccc}
\noalign{\smallskip}
\hline\noalign{\smallskip}
Isotope                & \multicolumn{3}{c}{Mass (mg)} \\
                       & 2 months             & 4 months            & 6 months \\
\hline\noalign{\smallskip}
$^{249}$Bk             & $7.5\cdot10^{-3}$   & 1.6                  & 32       \\
$^{249}$Cf             & $1.0\cdot10^{-4}$   & 0.042                & 1.2              \\
$^{250}$Cm             & $2\cdot10^{-7}$     & $1.1\cdot10^{-4}$   & $4.1\cdot10^{-3}$   \\
$^{251}$Cf             & $2\cdot10^{-6}$     & $8.2\cdot10^{-3}$   & 0.55                \\
$^{254}$Es             & $10^{-14}$           & $10^{-8}$           & $1.4\cdot10^{-5}$   \\
$^{257}$Es/$^{257}$Fm  & $-$                 & $3\cdot10^{-15}$     & $3\cdot10^{-11}$           \\
\hline\noalign{\smallskip}
\end{tabular}
\end{threeparttable}
\label{table:SHE_masses}
\end{table}

The yields of neutron-rich elements rapidly increase if the irradiation time becomes longer.
If the irradiation time is prolongated by a factor of 2, i.e. from six~months to one year,
the corresponding amounts of neutron-rich elements will increase by a factor of 
$$ F(n) = 2^n \frac
{{\rm e}^{-\bar{n}(1   \rm y)}}
{{\rm e}^{-\bar{n}(0.5 \rm y)}} 
 = 0.55 \cdot 2^n $$
where $n$ is the number of captured neutrons, see Eq.~(\ref{eq:poisson}). 
For $^{257}$Es $n=14$ and $F(14) \simeq 10^4$. According to the results of 
NuCoD calculations the mass of accumulated $^{257}$Es is $3\cdot10^{-11}$~mg,  
after half a year of irradiation, see Table~\ref{table:SHE_masses}. 
However, it was assumed that $^{257}$Es does 
not decay, while in reality it has a half-life time of about 7~days and 
undergoes $\beta^-$-decay into $^{257}$Fm, which has the half-life time of 
100~days. This means that calculated mass of $^{257}$Es actually corresponds to 
$^{257}$Fm. Multiplying $3\cdot10^{-11}$~mg by the factor of $F(14) \simeq 10^4$
one can obtain that the mass of accumulated $^{257}$Fm should be about 
$10^{-7}$~mg after one year of irradiation.

\section{Conclusions}
\label{sec:conclusion}

We have extended the MCADS model previously used for simulations of spallation 
targets irradiated by intensive proton and deuteron beams to account for changes 
of isotopic composition of the spallation target during a long-therm irradiation. 
For this purpose a dedicated software called 
Nuclide Composition Dynamics (NuCoD) was developed on the basis of MCADS. It 
models spallation targets taking into account the time evolution of their 
composition due to primary and secondary nuclear interactions and radioactive 
decays including nuclear reactions on spallation products. NuCoD was validated
against available experimental data~\cite{Bakhmutkin1986} on nuclide composition
in a Pb spallation target after its irradiation by a proton beam. 

NuCoD was used to study the possibility to produce transuranic neutron-rich 
super-heavy nuclei in spallation targets. It was found that macroscopic 
quantities of isotopes can be produced up to $^{249}$Bk. Therefore, ADS can be 
considered as an alternative for reactors currently used for the TRU elements 
production. It was demonstrated that $^{257}$Fm can be produced at least in 
picogram level with the state-of-the-art accelerator technology. Moreover, the 
production rate of SH nuclei can be increased with farther improvements of the 
system. For example, if the neutron capture rate would be enhanced by a factor 
of 4, e.g.\ by adding moderator materials, the abundance of $^{257}$Fm would 
increase by 7~orders of magnitude. Therefore, future development of accelerator 
technology together with optimization of the target assembly can open a new 
opportunity for producing neutron-reach heavy and super-heavy elements in 
macroscopic quantities.

\section*{Acknowledgments}

The authors thank Valeriy Zagrebaev and Alexander Karpov for fruitful discussions.
Our calculations were performed at the Center for Scientific Computing (CSC) of
the Goethe University, Frankfurt am Main. We are grateful to the staff of
the Center for assistance.

\section*{Appendix A -- Calculation of decay chains}
\label{app:d_chains}

The rate of the nuclide composition change $v(A,Z,n)$ between the steps $n+1$ and 
$n$ caused by primary and secondary particle-nucleus interactions can be 
calculated as:
\begin{equation}
\label{eq:v}
v(A,Z,n) = (a(A,Z,n+1) - a(A,Z,n)) \cdot I \cdot 6.241\cdot10^{15},
\end{equation}
where $I$ is the beam current in mA, which is multiplied by the factor of 
$6.241\cdot10^{15}$ to provide the beam current in particles per second.
Below the arguments $A$, $Z$ and $n$ are omitted for the sake of simplicity. 
Then the reduction of $a_0$, the abundance of an isotope in the target, due to 
its decay is defined by the formula:
\begin{equation}
\label{eq:a_0_variation}
\frac{{\rm d} a_0(t)}{{\rm d} t} = v_0 - \frac{a_0(t)}{\tau_0},
\end{equation}
where the index ``0'' is referred to the parent isotope in a decay chain and 
$\tau_0$ is its mean lifetime. The solution of Eq.~(\ref{eq:a_0_variation}) 
is:
\begin{equation}
\label{eq:a_0}
\begin{array}{l}
a_0(t) = b_0 + c^0_0 e^{-t/\tau_0} \\
b_0 = v_0 \tau_0 \\
c^0_0 = a_0(0) - v_0 \tau_0. \\
\end{array}
\end{equation}
The abundance variation of the direct product of the parent isotope with the 
abundance $a_0$ is defined by the formula:
\begin{equation}
\label{eq:a_1_variation}
\frac{{\rm d} a_1(t)}{{\rm d} t} = \frac{a_0(t)}{\tau_0} - \frac{a_1(t)}{\tau_1},
\end{equation}
where $\tau_1$ is the mean lifetime of the daughter isotope. The solution of this 
equation is:
\begin{equation}
\label{eq:a_1}
\begin{array}{l}
a_1(t) = b_1 + c^0_1 e^{-t/\tau_0} + c^1_1 e^{-t/\tau_1} \\
b_1 = v_0 \tau_1  \\
c^0_1 = ( a_0(0) - v_0 \tau_0 ) \frac{\tau_1}{\tau_0 - \tau_1} \\
c^1_1 = - v_0 \tau_1 - ( a_0(0) -v_0 \tau_0 ) \frac{\tau_1}{\tau_0 - \tau_1}. \\
\end{array}
\end{equation}
For the subsequent isotopes in the decay chain:
\begin{equation}
\label{eq:a_i_variation}
\frac{{\rm d} a_i(t)}{{\rm d} t} = \frac{a_{i-1}(t)}{\tau_{i-1}} - \frac{a_i(t)}{\tau_i}.
\end{equation}
The solutions can be obtained recurrently:
\begin{equation}
\label{eq:a_i}
\begin{array}{l}
a_i(t) = b_i + \sum^{i}_{k=0} c^k_i e^{-t/\tau_k} \\
b_i = b_{i-1} \frac{\tau_i}{\tau_{i-1}} \\
c^k_i = c^k_{i-1} \frac{\tau_i \tau_k}{\tau_{i-1}(\tau_k-\tau_i)},\; k \neq i \\
c^i_i = - b_i - \sum^{i-1}_{k=0}c^k_i. \\
\end{array}
\end{equation}
A simple Python-code calculates abundance variation rates $v(A,Z,n)$ for all 
the isotopes using the output of MCADS simulation with the base target and the 
isotope-enriched targets according to Eqs.~(\ref{eq:a_n+1}) and (\ref{eq:v}). 
Then the code runs through all the isotopes, either existed in the primary 
target at the beginning of the step or with a non-zero value of $v(A,Z,n)$, and 
recalculates the corresponding abundance according to Eq.~(\ref{eq:a_0}). For 
stable isotopes $\tau = \infty$ is taken. After that, the abundance of all 
the subsequent decay products are calculated according to Eq.~(\ref{eq:a_i}).

\section*{References}
\bibliography{ADS_Literature,Americium}
\end{document}